\documentclass{article}

\usepackage[T1]{fontenc}
\usepackage[utf8]{inputenc}
\usepackage{natbib}
\usepackage{color}
\usepackage{booktabs}
\usepackage{tabulary}
\usepackage{siunitx}
\usepackage{tikz}
\usepackage{graphicx}
\usepackage{latexsym}
\usepackage{subfigure}
\usepackage{multirow}
\usepackage{verbatim}
\usepackage{moreverb}
\usepackage{listings}
\usepackage{float}
\usepackage{graphics}
\usepackage{paralist}
\usepackage{amssymb}
\usepackage{amsmath}
\usepackage{listings}
\usepackage{lscape}
\usepackage{wrapfig}
\usepackage{lstautogobble}
\usepackage{enumerate}
\usepackage{url}
\usepackage{textcomp}
\usepackage{todonotes}

\usepackage{authblk}

\begin{document}

\title{Capturing Knowledge of Emerging Entities From Extended Search Snippets}

\author[1]{Sunday C. Ngwobia}
\author[1]{Saeedeh Shekarpour}
\author[2]{Faisal Alshargi}

\affil[1]{University of Dayton, Dayton, USA}
\affil[2]{University of Leipzig, Leipzig, Germany}

\maketitle

\begin{abstract}
 Google and other search engines feature the entity search by representing a knowledge card summarizing related facts about the user-supplied entity.
 However, the knowledge card is limited to certain entities that have a Wiki page or an entry in encyclopedias such as Freebase. The current encyclopedias are limited to highly popular entities, which are far fewer compared with the emerging entities.
Despite the availability of knowledge about the emerging entities on the search results, yet there are no approaches to capture, abstract, summerize, fuse, and validate fragmented pieces of knowledge about them. 
Thus, in this paper, we develop approaches to capture two types of knowledge about the emerging entities from a corpus extended from top-n search snippets of a given emerging entity. The first kind of knowledge identifies the role(s) of the emerging entity as, e.g., who is s/he? The second kind captures the entities closely associated with the emerging entity. 
As the testbed, we considered a collection of 20 emerging entities and 20 popular entities as the ground truth. Our approach is an unsupervised approach based on text analysis and entity embeddings. Our experimental studies show promising results as the accuracy of more than $87\%$ for recognizing entities and $75\%$ for ranking them. Besides $87\%$ of the entailed types were recognizable. 
Our testbed and source code is available on Github  \url{https://github.com/sunnyUD/research\_source\_code}.
\end{abstract}



\section{Introduction}
Google coined the \textbf{knowledge card} in 2012, which is a small panel on the right side of the search results obtained when a end-user looks or search for a Named Entity (NE) -throughout this paper, we will refer NE as an entity. 
For example, when the user inquires about \texttt{Barack Obama}, a knowledge card is represented on the right side.
This card displays a summarized version of significant facts about the target entity.
However, the representation of knowledge card is limited to popular entities, meaning only those entities which have an entry in encyclopedias such as Wikipedia.
Nevertheless, a large portion of target entities within the search queries might not have any Wiki page because they might be emerging or not adequately well-known or popular for publishing in a Wiki page. 
Although search engines such as Google do not show any knowledge card, the search results contain precious fragments of knowledge about emerging entities. 
We call such entities as \textbf{emerging} or unpopular entities.
For example, if the target entity is \texttt{Saeedeh Shekarpour} who is a researcher, the search results do not demonstrate a knowledge card.
Despite this limitation, the top search results retrieved for the target entity contain precious and relevant knowledge scattered across various documents. However, the end-user has to navigate and explore these documents one by one to acquire reasonable background knowledge about the target emerging entity. 
In this paper, our research question is concerned with developing novel methodologies that \textbf{capture the knowledge of emerging (unpopular) entities from the top-n search snippets}.
Although there is a body of research for knowledge extraction from Web pages \cite{zhao2017web,hoffmann2011knowledge,pasupat2014zero}, however, they are mainly limited to capturing knowledge of popular entities \cite{kazama2007exploiting,zaghloul2017developing}. 
%
We categorize three types of knowledge entailment for the emerging entities as follows:
(i) \textbf{Type of entity:} refers to the types that the target entity can be attributed, e.g., the entity \texttt{Saeedeh Shekarpour} holds the type of \texttt{researcher} or \texttt{Assistant Professor}. (ii) \textbf{Associated Entities:} engages entities that hold a relationship with the target entity. For example, the entity \texttt{University of Bonn}, which is an organization, is associated with \texttt{Saeedeh Shekarpour} where she obtained her Ph.D. (iii) \textbf{Associated facts:} represents the certain facts such as \texttt{birth date} about the target entity. E.g., \texttt{Semantic Web} is related to the entity \texttt{Saeedeh Shekarpour} with respect to the relation \texttt{Research Area}. 
These three types of knowledge can be leveraged to shape a entity-centric knowledge graph about the target emerging entity.
In this article, we contribute to the following areas (i) developing an approach to entail the entities associated with a given emerging entity, (ii) developing an approach to entail the type of the emerging entity, (iii) proposing an approach that compiles a textual corpus from search snippets concerning the rank of content, (iv)developing a testbed for evaluation purposes.
This paper is organized as follows: Section 2 reviews the related work.
Section 3 states and formalizes the concerning problem. Then, Section 4 presents our approach, followed by our experimental study in Section 5.
We close with the remarks of conclusion and future plan.

\section{Related Work}
\label{sec:related-work}

Knowledge Graphs (KGs) are becoming a crucial resource \cite{wang2018acekg} to support various AI-based applications such as graph analytics and question answering systems.
However, the state-of-the-art of KGs are incomplete(i.e full coverage), and the KG completion is a task of high importance \cite{Paulheim2017KnowledgeGR}.
The task of capturing new knowledge can be divided into two stages as first recognizing the emerging entities and then capturing the knowledge about the emerging entities.
Our paper contributes to capturing knowledge about the emerging entities. The assumption is that the search query contains a surface form of an emerging entity.
The work presented in \cite{hoffart2014discovering} describes emerging entities such as new songs, movies, companies, people where they are out-of-knowledge-base (OOKB) entities. Their work was basically discovering emerging entities in a text document by harvesting all the key-phrases for all mentions in the document and using those to run a co-occurrence count on the YAGO (Yet Another Great Ontology) knowledge graph to determine if it is an emerging entity. But our approach is an extension of this because it captures knowledge about a target emerging entity which is already recognized (by the end-user).
Being OOKB causes loss of recall in the majority of Named Entity Recognition (NER) tools.
For example, the paper \cite{derczynski2017results} states that most of the unsupervised entity recognition approaches claim a high recall and precision scores for popular entities, whereas they do not perform well with rare or emerging entities due to the variations in their surface forms. 
The work presented in the article \cite{lilleberg2015support} describes a named entity as a sequence of words referring real-world entities such as \textit{Google Inc}, \textit{United States}, and \textit{Barack Obama}. The task of named entity recognition is to identify entities in a free text and classify them into any of the predefined categories such as person, location, organization, etc. Further, it learns about the relationship between these entities with a machine or deep learning models through the entities co-occurrence matrix or semantic vector. 
 \cite{zaghloul2017developing} suggests how to create high precision seed list to improve extraction of salient named entities from a corpus. Furthermore, according to \cite{remmiya2018co}, word embedding models could be used to retrieve the vector representation of each word from co-occurrences of context word. We leveraged this concept in extraction of associated entities of a target emerging entity. Also, we used their concept of creating a high precision seed list in conjunction with Stanford-NER to extract the coarse type entity from our extended and enhanced corpora that we used in the training entity embedding model.
Obtaining the entity type is another crucial stage in knowledge capturing tasks. \cite{del2015finet}, describes entity typing as the task of detecting the type(s) of a named entity in context. They used a supervised method that performs entity type extraction, and among the various types of extractor they used in their experiment, is the corpus-based extractor. Their method employs the POS tagging and WordNet annotation. Also \cite{yaghoobzadeh2018corpus} used a CNN(Convolutional Neural Network) embedding model with distant supervision to perform entity typing. However, our approach performs corpus-based entity type extraction using word2vec embedding and WordNet framework.
Another way to achieve this task of entity typing is by considering the relational pattern between an entity and its neighboring entities. Patterns for relations are a vital ingredient for many applications, including information extraction and question answering. If a large-scale resource of relational patterns were available, this could boost progress in NLP and AI tasks \cite{nakashole2012patty}. So, their approach involves using syntactic, ontology and lexical(SOL) pattern to entail entity types which is different from our approach of using tf-idf method and wordNet synset to entail entity type. Their approach precision score was 84.7\% while our approach precision score is 87\% precision. Therefore, our approach performs better thus can be integrated into tasks such as  knowledge graphs types enrichment, relation extraction \cite{yaghoobzadeh2016noise}, question and answering \cite{yavuz2016improving}, search and query analysis \cite{balog2012hierarchical} and co-reference resolution \cite{durrett2014joint}. 
Regarding the fine-grained entity typing task, the paper \cite{choi2018ultra}  introduces the head-word distant supervision approach to predict free-form noun phrases as types that the target entity plays in a given sentence. We used this noun-phrase approach in entailing the types of the target emerging entities. Whereas \cite{fan2014distant} describes the use of distantly supervised relation extraction approach as an incomplete multi-label classification problem with sparse and noisy features.  Therefore, they prefer solving the classification problem by using matrix completion on the factorized matrix of minimized rank.
Generally, we noticed that most of the research papers focus on popular entities with least attention to emerging or unpopular entities despite the enormous data available about them in the Internet, it is challenging to carry research evaluation because the datasets are not labelled. Our observation of this lapses motivated us to develop a model for extracting knowledge of entities being unpopular, however, emerging from the extended top Google snippets.



\section{Problem Statement}

We aim at capturing the knowledge of a given entity (denoted by $e_t$) supplied by the end-user from the extended top-n search snippets. \textit{Please note that a given snippet is extended by including the full content to which the snippet link points out. In other words, extended top-n snippets are the extract snippet plus the textual content of the Html page to which the link in the snippets points}.

The first kind of captured knowledge indicates the corresponding type of the given entity using \texttt{rdf:type} (i.e., \url{https://www.w3.org/TR/rdf-schema\#ch\_type}) relation from Resource Description Framework.
This relation assigns a particular type described in the class $C_i$ to the given entity.
Thus, the captured knowledge is represented as the triple \texttt{$(e_t,$\texttt{rdf:type}$,C_i)$ }which associates a particular type to the target entity $e_t$.
Regarding our running example, Figure \ref{fig:entity_exm} shows the transformation process of the extended top-n snippets to the knowledge graph for the given entity.
For example, since the target entity is \texttt{Saeedeh Shekarpour}, as illustrated in Figure \ref{fig:entity_exm}, three types (nodes with orange color) are entailed, which can be serialized using turtle format in the following triples: \texttt{Saeedeh-Shekarpour rdf:type    Researcher, \hspace{5pt}Scholar, \hspace{10pt}Assistant  Professor.}
The second kind of knowledge implies the entities associated with the given entity.
Such associations require inferring two pieces of knowledge as the first piece entails the associated entity and the second piece entails the type of the relation which associates the two entities.
For example, regarding the example of Figure \ref{fig:entity_exm}, the entities such as \texttt{Germany}, \texttt{S{\"o}ren Auer} are entailed.
In this paper, we do not deal with entailing specific types for relations and bound to generic relations. For example, for the relationships between entities of the type \texttt{Person}, we use the generic relation \texttt{foaf:knows}.
Another example is the generic relation \texttt{https://schema.org/location} for associating a location to a given entity.
Thus, formally the associated entities are represented as the triple $(e_t,rel,e')$ which associates a particular entity $e'$ to the given emerging entity $e_t$.

\paragraph{\textbf{Aim.}} We aim at entailing the two sets as follows: (i) the set of all ontological classes (concepts) $C=\{C_1,...,C_i\}$ that can be assigned as the type of a given entity $e_t$. (ii) the set of all entities $E=\{e_1,...,e_j\}$ that can be associated with the given entity $e_t$. 

\section{Approach}
\label{sec:approach}
We illustrate a top overview of our approach in Figure \ref{fig:architecture}.
Our approach relies on a corpus that is collected on-the-fly for an entity supplied by the search query of the end-user.
Thus, a given input entity is sent to Google API i.e., \url{https://github.com/ecoron/SerpScrap}.
Then, a corpus of snippets is compiled from the first eight snippets of the Google search results.
The next step is the corpus extension, which includes the textual content of the page, which the snippet refers to.
The extended corpus is a contextual corpus because it contains more abundant contextual information about the target entity. This corpus is based on entailing type and associated entities of the target entity.
In the following subsections, we describe the details of our approach for capturing the knowledge of a given target entity.
\subsection{Entailing Associated Entities for Emerging Entity}
\par The entailment of associated entities implies extracting entities from the extended corpus where they hold a high chance of being related to the given entity.
To compute the degree of relatedness, we rely on the embeddings trained for entities.
In other words, we acquire the embeddings for entities from a model trained on our extended corpus (or a variation of our corpus, which will be discussed in the following). Then, the cosine similarity of the vector of each entity with the vector of the target entity is computed.
The higher cosine similarity, the higher relatedness bound.
To accomplish that, we follow the four steps listed below, and it is a synopsis of the functions of the various units represented in our model in figure 1:
\vspace{0.1cm}\newline
\textbf{Step 1: Corpus Pre-processing.}
This step normalizes the underlying corpus to increase the performance for tasks such as Named Entity Recognition (NER). For example, it removes special characters, digits, and stop-words. It consists of the \texttt{serScrap} and \texttt{snippet collation} modules in figure 1.\vspace{0.1cm} \\
\textbf{Step 2: Named Entity Recognition.}
In this step, the entities with the primary types (i.e., person, organization and location) are recognized from the underlying corpus using Stanford core-NLP Named Entity Recognizer (NER) tool \cite{manning2014stanford}.\vspace{0.1cm}\\
\textbf{Step 3: Entity Ranking using Embeddings (entity2vec)}
The goal of this step is to identify the entities which are highly associated with the given target entity. To identify them, we rely on the cosine similarity between the given emerging entity and entities recognized from the previous step. To compute that, we have to train the embedding models (e.g., word2vec embedding model \cite{word2vec1,word2vec2} ) for entities. 
Then the vector of entities is used to compute the degree of relatedness of entities within the corpus to the given entity. 
We have to transform the recognized entities into a single phrase representation in order to be able to train a word-level embedding model such as word2vec for entities.
Thus, we concatenate words of a multi-word entity into a single-word phrase using the \# symbol. For example, the entity \texttt{Saeedeh Shekarpour} is concatenated as \texttt{Saeedeh\#Shekarpour}. 
After transforming the underlying corpus using the new representation of entities (by replacing the previous representation of entities into the new ones), we employ the skip-gram model of the word2vec approach \cite{word2vec1,word2vec2} to acquire the embedding of entities. \vspace{0.15cm}\\
\noindent\textbf{Rank2Corpus.}
We contribute to a novel approach that reflects the rank of the search results into the corpus and subsequently in the embedding model. 
The rationale behind our approach (\texttt{Rank2Corpus}) is to prioritize the information (e.g., entities) based on which appear in the higher positions of the search results. By doing so, we aim to build a new corpus that will boost the score of relevant associated entities contained in each retrieved google snippets for a given emerging entity search result. The new corpus takes into account the ranking of snippets from Google search results, and we refer to it as an enhanced corpus.
In essence, the content from a higher position (rank) is more relevant than the content from a lower position (rank).
Given the above fact, we increase the frequency of content acquired from each snippet using the equation 
$ corpus^* = \bigcup_{1}^{n} \bigcup_{1}^{n+1-i} S_i$. 
where $corpus^*$ denotes the enhanced corpus (new corpus) which takes into account the rank of the content, $S_i$ is the content of the extended snippet positioned in the $i$the place of the search results and $n$ refers to the top-n search results included for developing the corpus. 
We refer to this corpus as the \textbf{enhanced corpus} $corpus^*$, where it includes the rank by adding the frequency.
The other corpus which does not consider ranking is called the extended corpus $corpus^+$. In our experiments, we employ both of these corpora for comparison purposes. 
Thus, for example, if we obtain the top-10 search snippets, the content from the snippet in position one is repeated ten times in the enhanced corpus, whereas the content from the position ten is repeated only once.
In fact, this approach increases the statistics of the content within the corpus based on the rank of its snippet.
The higher co-occurrence of text eventually leads to a higher similarity score.
The embedding models are typically based on the statistical approach (they count frequency of the context), and in addition, they require to be trained on a sufficient volume of a corpus in order to attain optimal performance.
In our case, the models are expected to capture salient entities that are highly associated with the given emerging entity. 
Thus, once we increase the frequency of content based on their rank, it provides more context for the entities occurring in the higher rank and suppresses the entities from the lower rank.
Thus, unevenly enhancing the frequency of content implies to capture the ranking inside the corpus.
In the case of having $n$ snippets, if the content comes from the rank $i$, then its content is repeated in the corpus for $n+1-i$ times.
In other words, the equation connotes that snippets are reproduced \textit {n times} based on their ranking position. So, the first snippet $ S_1$ with the highest rank will be replicate \textit{n+1-1} times, the second $ S_2 $ for \textit{n+1-2} times, ..., and the last $ S_n $ for \textit{n+1-n} times \vspace{0.03cm}.



\begin{figure}[hptb]
\centering
 \subfigure[On the extended corpus: $corpus^+$. Here, there is no clear separation of entities based on type.]{
   \includegraphics[width=0.45\columnwidth, height=5cm ]{ 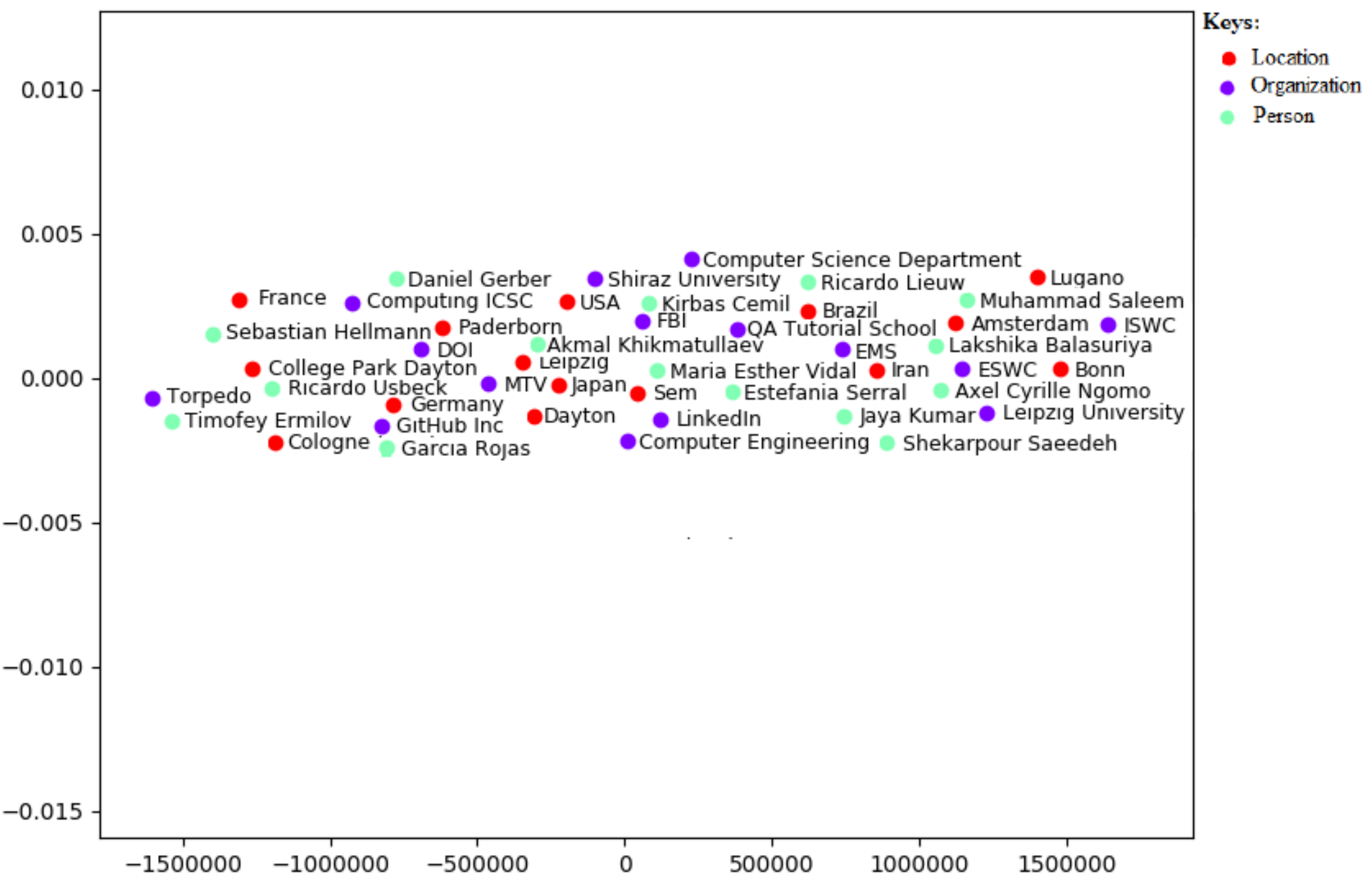}
   \label{fig:pca_Saeedeh_a}
	}
 \subfigure[On the enhanced corpus: $corpus^*$. Here, there is a clear separation of entities based on type.]{
  \includegraphics[width=0.45\columnwidth,height=5cm]{ 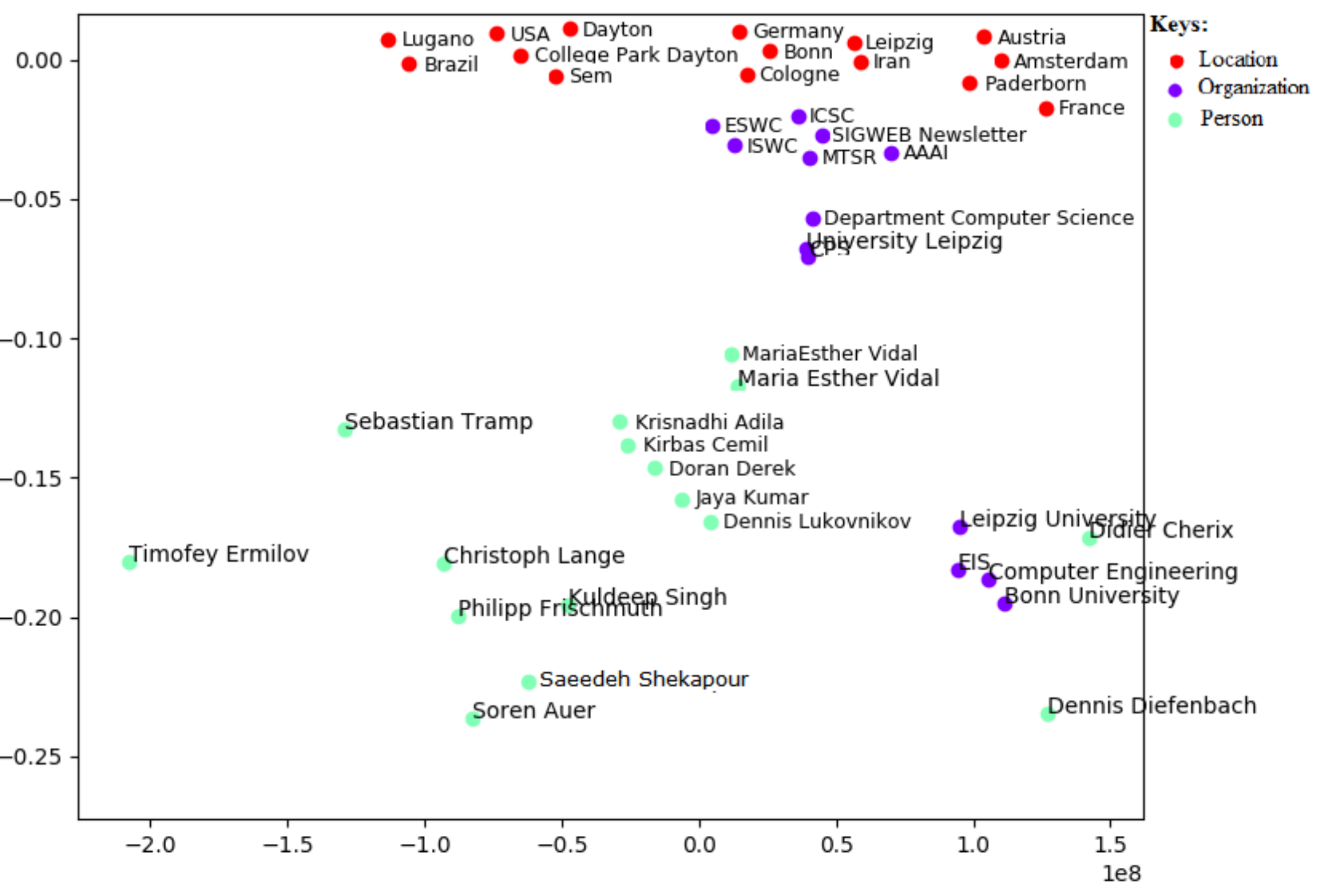}
 \label{fig:pca_Saeedeh_b}
   }
 \caption{PCA results for the emerging entity \texttt{Saaedeh Shekarpour with the type \texttt{Person}.}}
  \label{fig:pca_Saeedeh}
 \subfigure[On the extended corpus: $corpus^+$. Here, there is no clear separation of entities based on type. ]{
   \includegraphics[width=0.45\columnwidth,height=5cm]{ 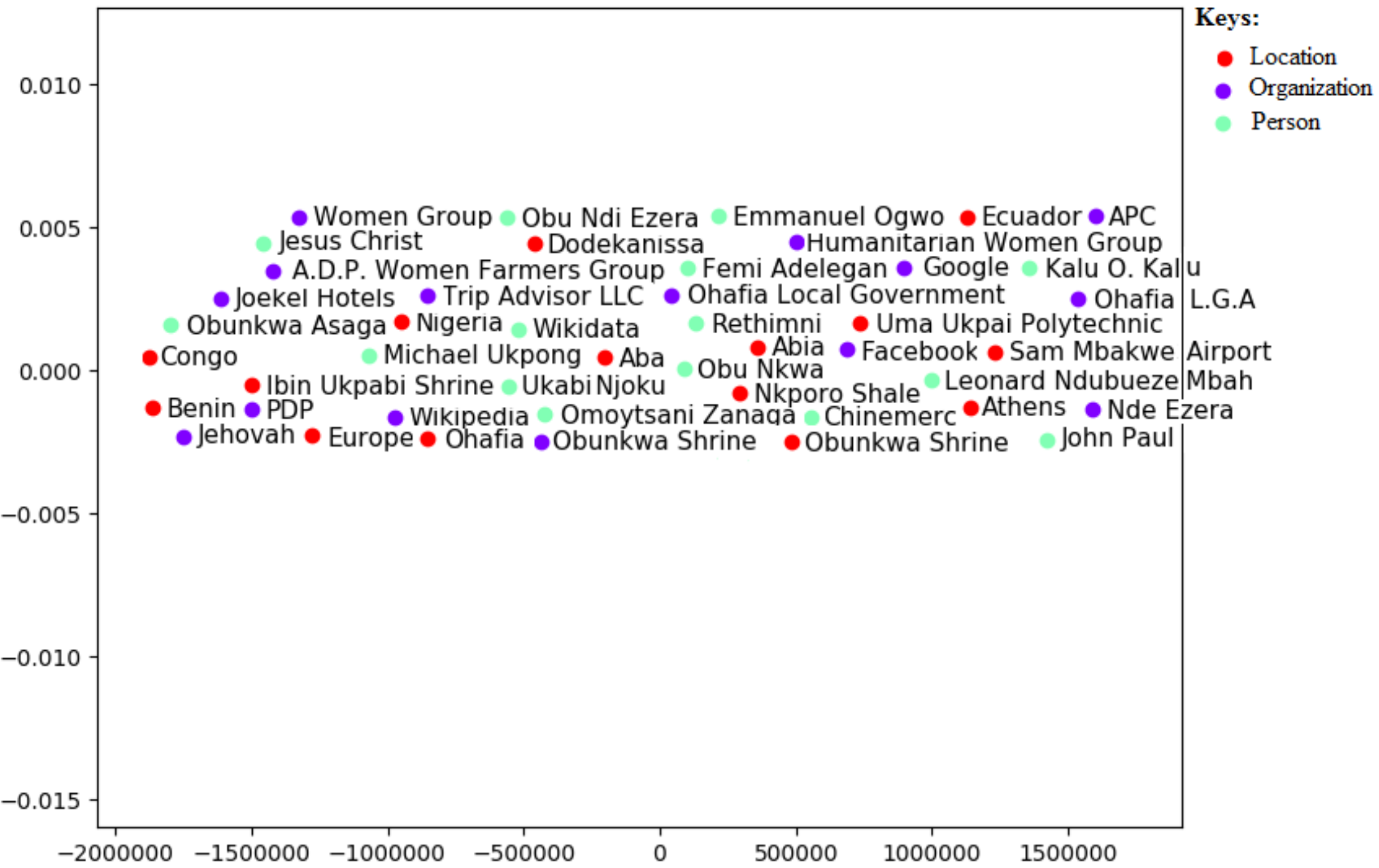}
   \label{fig:pca_asaga_1a}
	}
 \subfigure[On the enhanced corpus: $corpus^*$. Here, there is a clear separation of entities based on type. ]{
   \includegraphics[width=0.45\columnwidth,height=5cm]{ 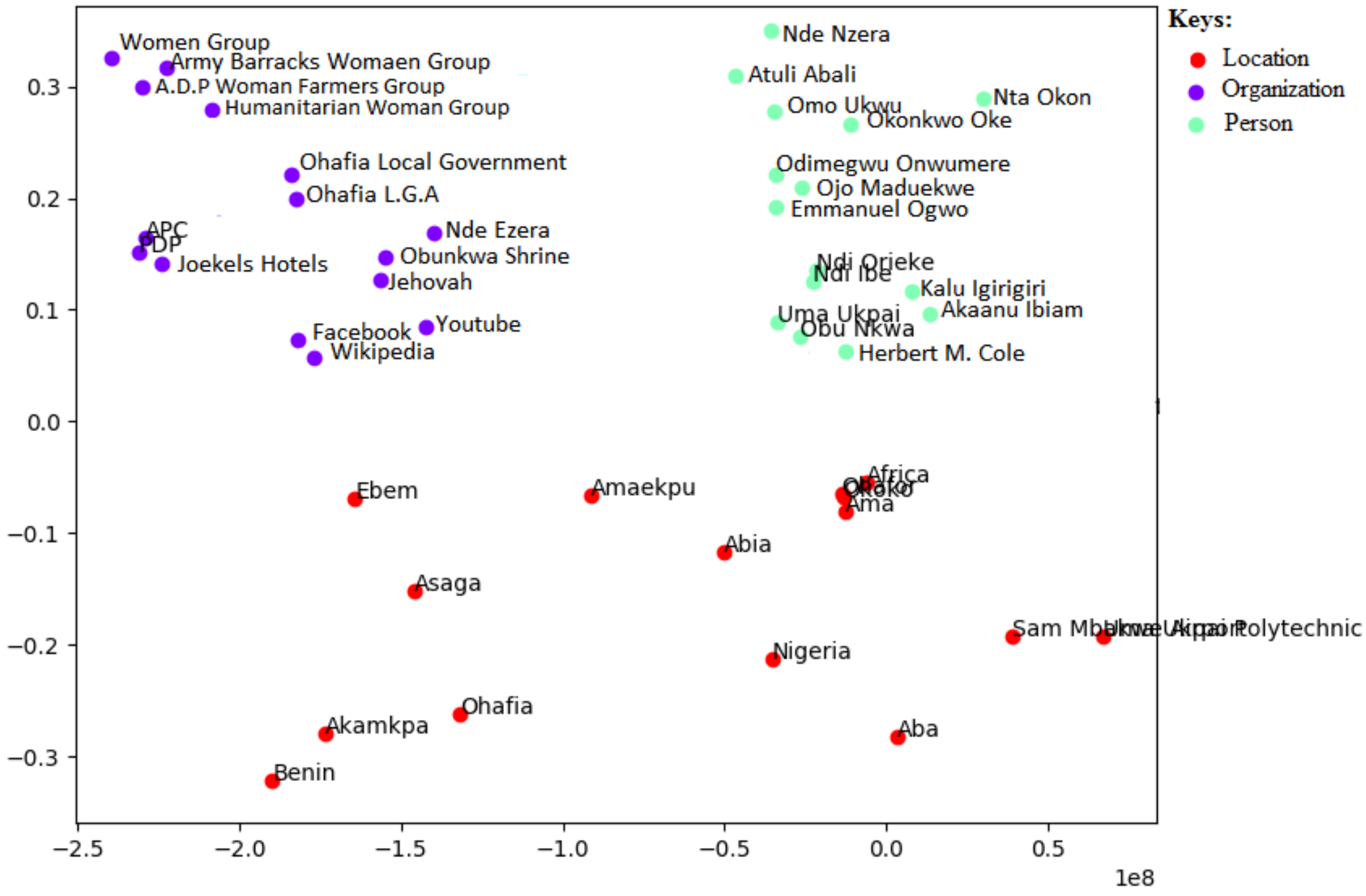}
    \label{fig:pca_asaga_1b}
   }
 \caption{PCA results for the emerging entity \texttt{Asaga Ohafia in Nigeria} with the type \texttt{Location}.}
  \label{fig:pca_asaga_ohafia}
 
 \subfigure[On the extended corpus: $corpus^+$. Here, there is no clear separation of entities based on type. ]{
   \includegraphics[width=0.45\columnwidth,height=5cm]{ 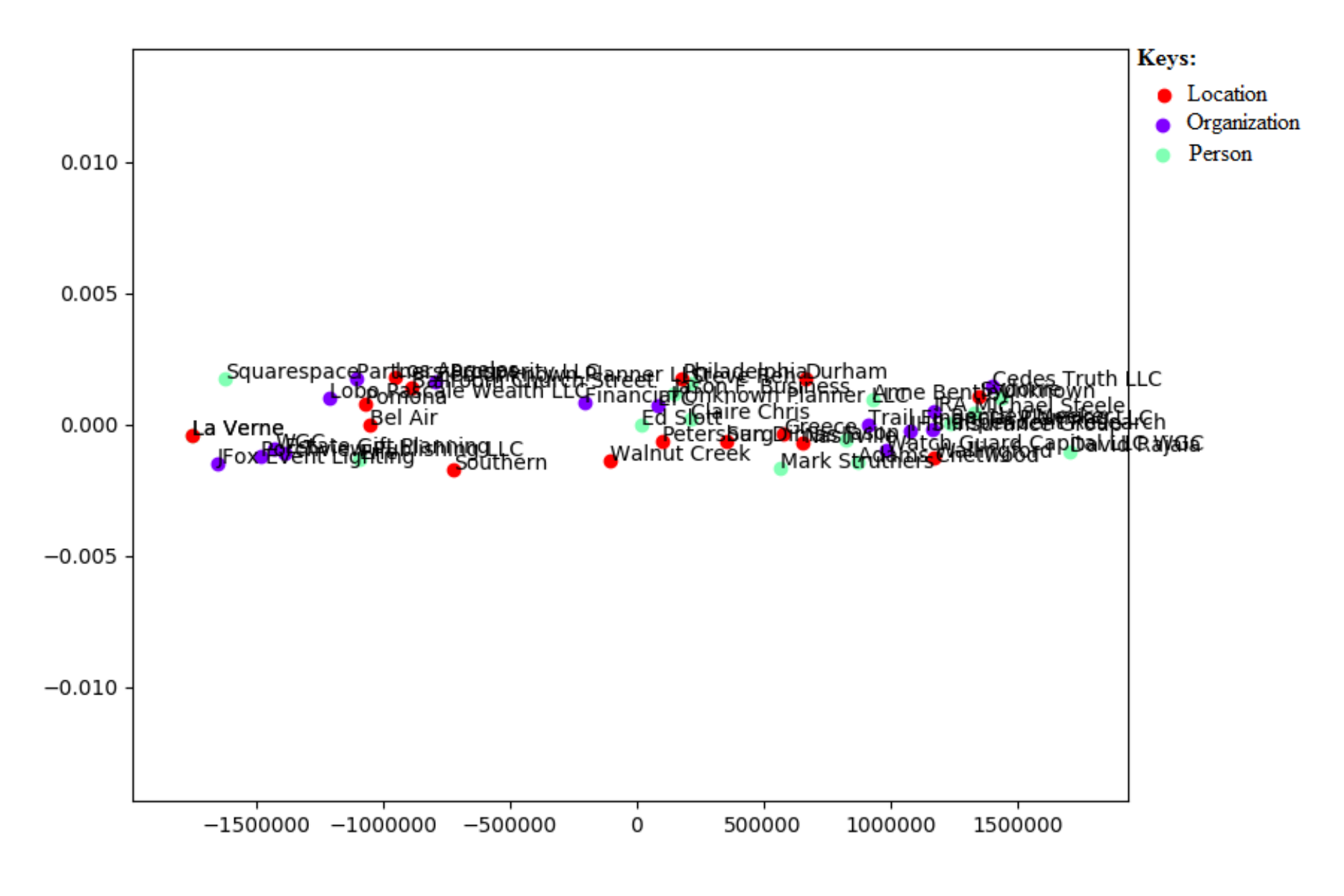}
   \label{fig:pca_unknown_planner_1a}
	}
 \subfigure[On the enhanced corpus: $corpus^*$. Here, there is a clear separation of entities based on type. ]{
   \includegraphics[width=0.45\columnwidth,height=5cm]{ 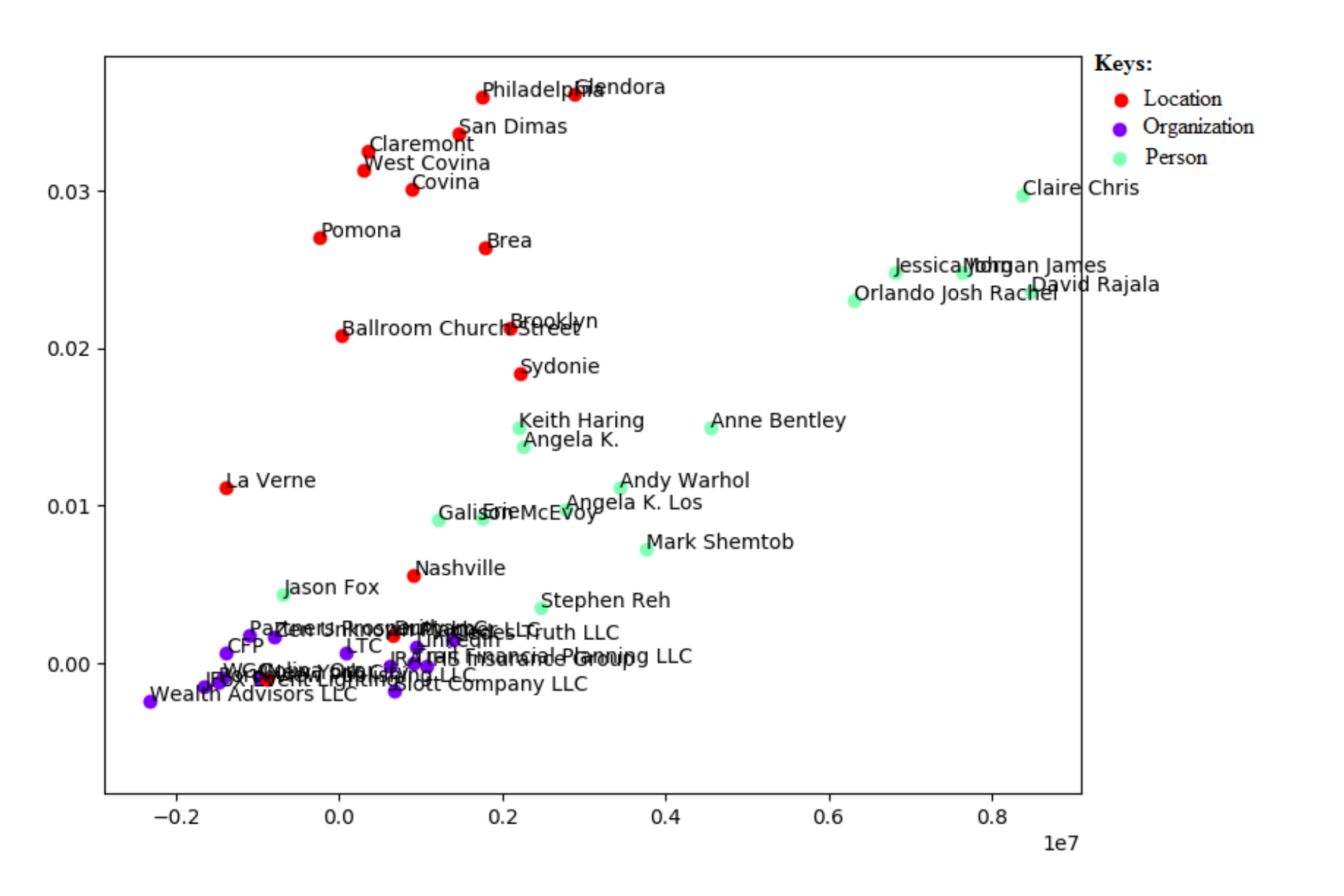}
    \label{fig:pca_unknown_planner_1b}
   }
 \caption{PCA results for the emerging entity \texttt{Unknown Planner LLC} with the type \texttt{Organization}.}
  \label{fig:pca_unknown_planner}
\end{figure}

\noindent \textbf{Step 4: Visualization and Clustering using PCA.}
The work presented in \cite{metrics} introduces several metrics for measuring the quality of embeddings, particularly for the knowledge graph. 
One of the metrics is using a visualization approach to make insight over the behavior of the embeddings in the 2-dimensional space. Thus, herein, we rely on the visualization and clustering approaches over the vectors learned for the entities to observe the behavior of the embeddings. 
We plot the result from our model using Principal Component Analysis (PCA) \cite{molino2019parallax} for a visual representation of the correlation between an emerging entity and associated entities as well as for analysis purposes at the later part of this work. 

\vspace{0.15cm} \noindent \textbf{ Disambiguation Challenge.} 
\cite{zhang2018approach} is a disambiguation approach relied on context similarity and entity correlation. Similarly, we adopt the context similarity method because it provides abstract entity information that aids the disambiguation of an entity. Thus, we leverage the word2vec model to implement it.
Since the target emerging entities have no wiki-page or knowledge graph record, we are not able to apply methods used in \cite{parravicini2019fast} for disambiguating popular entities.
Still, disambiguation is an open research area for our future work.
\subsection{Entailing Type for Emerging Entity}
Here, we are not talking about the coarse-grained (\textit{e.g person, location, organization, etc.}) types for the targeted emerging entities but of their fine-grained types (\textit{e.g professor, footballer, researcher, grocery store, etc.}).
Our type entailment approach is inspired by the paper \cite{choi2018ultra}
where it considers the head-word distant supervision approach to predict free-form noun phrases as types.
We rely on a modified and unsupervised version described in the following.
The entailment of the type of emerging entities relies on text processing over the extended corpus. In principle, it contains three stages listed as follows: (i) Pruning of Named Entities (PNE), (ii) Extraction of Noun Phrases (NP), (iii) Term Frequency analysis of the extracted NPs.
The named entities recognized in the previous phase are pruned from the extended corpus.
Because they do not contribute to determining types.
Thus, all of them are removed. 
Next, the nouns and noun phrases (which are called terms in this context) are driven from the extended corpus $corpus^+$ utilizing WordNet framework.
Finally, the popularity or relatedness of the all driven terms is computed via a \textbf{modified} term-frequency inverse-document-frequency (tf-idf) technique.
Such computation is formalized in the following equations: $tf_t=\sum_{d\in D} log(1 + f_{t,d})$, $idf_{t}=\frac {N}{df_t}$, and $tf-idf_t=tf_t\times idf_t$. 
\\
Where d specifies a document (content acquired from one single snippet) in the corpus, and D shows the collection of documents. Further, $t$ denotes the given term and $f_{t,d}$ refers to the raw frequency of the term $t$ within the document $d$.
Equation 2 computes the sum of the logarithmic term frequency of $t$ over all the documents. 
Equation 3 computes the inverse document frequency for the term $t$ where $N$ is the total number of documents in the collection $D$, and $df_t$ is
the number of documents containing the term $t$.
The ultimate score $tf-idf$ for every term $t$ is computed in Equation 4 by multiplying the two previous scores.
In our case, we choose the top m (m=2 or 3 in our work) terms with the highest $tf-idf$, and then they are considered as the type for the emerging entity.
Furthermore, Table 1 illustrates how this approach works using two samples of the snippet. The first column highlights the named entities in the snippet. The highlighted text will be removed together with stop-words, as indicated in the second column. 
The third is the result obtained by extracting the noun phrases (NPs) in column 2. The final column consists of the first top 2 or 3 terms with the highest $tf-idf$ score. And this will be used in the entity typing.
\begin{table}
\begin{scriptsize}
\caption{ The demonstration of NP extraction using example.}
\begin{tabular}
{>{\raggedright\arraybackslash} m{5.5cm}|m{2.8cm}|m{2cm}|m{1.5cm} }
 
\toprule
 Snippet &  Removal of stop-words and NE &  NP Extraction &  $tf-idf$ process  \\ 
\midrule
\colorbox{yellow}{ \textcolor{red}{Saeedeh Shekarpour}} Assistant Professor \colorbox{yellow}{ \textcolor{red}{Department}} of \colorbox{yellow}{ \textcolor{red}{Computer Science University}} of \colorbox{yellow}{ \textcolor{red}{Dayton}} News and Opportunities am founding \colorbox{yellow}{ \textcolor{red}{CANAB: Cognitive ANalytics}} \colorbox{yellow}{ \textcolor{red}{Lab}} in the \colorbox{yellow}{ \textcolor{red}{University}} of \colorbox{yellow}{ \textcolor{red}{Dayton}}, looking for talented, hardworking and passionate students.
 &  Assistant Professor News opportunities founding looking talented hardworking passionate student
 &   Assistant Professor \hspace{1.5cm}
News \hspace{1.5cm}
Students 
 & Assistant
 \hspace{1.5cm}
 Professor \hspace{1.7cm}
News\hspace{1.6cm}
Students
 \\

\bottomrule

\end{tabular}

\end{scriptsize}


\end{table}

\section{Experimental Study}
\label{sec:experiemnt}
In this section, we run a series of experimental studies to observe the effectiveness of our approach in capturing the knowledge of the emerging entities. 
Initially, we present our underlying testbed, followed by experimental settings. Next, the experimental studies respectively for entailing knowledge of the associated entities and types, are discussed. \vspace{-0.3cm}\newline

\noindent \textbf{Testbed:}
We require entity-centric corpus; thus, for any given entity, we have to compile its corresponding corpus.
Thus, we have a collection of entity-dependent corpora.
To compile our corpora, in an on-the-fly manner, for a given entity, we compile its corpus as described earlier.
As the testbed, we consider a collection of emerging entities based on coarse types as follows, ten entities of type \textit{Person}, five entities of type \textit{City}, and five entities of type \textit{Organization}. 
These entities were deliberately selected as they do not have any knowledge cards. However, there are informative snippets about them. 
In addition to this collection of emerging entities,
we considered another collection containing popular entities that Google shows a knowledge card for them. This second collection plays the role of ground truth.
The statistics of the second collection are similar to the first as it contains ten entities with the type  \textit{Person}, five entities with the type \textit{City} and five entities with the type \textit{Organization}.
So, in total, we have a collection of 40 entities where half of them are emerging, and the other half are popular entities.\vspace{0.15cm} \\
\noindent \textbf{Experimental Settings.}
We used skip-gram model of the word2vec embedding algorithm \cite{word2vec1,word2vec2} from the Gensim library\footnote{\url{https://radimrehurek.com/gensim/}} to implement our entity2vec embedding model. The tuned hyper parameters of our embedding skip-gram model are:
Window size = 7, Min-count = 1, Workers= 4, Size= 280, Batch word= 300, and sg= 1.\vspace{0.15cm} \\
\noindent \textbf{Experimental Study of Entailing Associated Entities}
To ascertain the effectiveness of our approach, we rely on three major methods of evaluation as follows: (i) Visualization of the associated entities for a given emerging entity; (ii) Human judge over the top-n associated entities of the emerging entities; (iii) Comparison of the associated entities of the popular entities with the Google Knowledge Card. \vspace{0.15cm} \\
\noindent \textbf{Top-k Associated Entity.} We separately trained the embedding models over the extended corpus $(corpus^+)$ and the enhanced corpus $(corpus^*)$ for each  target emerging entity in the set of 40 emerging entities that constitute our testbed.
Then, from each corpus, the top-k associated entities (entities with the highest cosine similarity to a given emerging entity as obtained from our model) are drawn.\vspace{0.15cm}

\noindent \textbf{Visualization.}
We represent the PCA results for each emerging entity. 
Herein, we represent a single sample for three types (person, location, and organization) of entities as follows: Figure \ref{fig:pca_Saeedeh} is the PCA result over the associated entities of the emerging entity \texttt{Saeedeh Shekarpour} with the type \texttt{Person}.
Figure \ref{fig:pca_asaga_ohafia} is the PCA result of the associated entities of the emerging entity \texttt{Asaga Ohafia}  (a place in  Nigeria) with the type \texttt{Location}. Finally, Figure \ref{fig:pca_unknown_planner} is the PCA result of the associated entities for the emerging entity \texttt{Unknown Planner LLC} with the type \texttt{Organization}.
In all of these three figures, the sub-figures labelled "a" are obtained from the extended corpus $(corpus^+)$ and the sub-figures labelled "b" are obtained from the enhanced corpus $(corpus^*)$.
As it can be observed, the  sub-figures labelled "b" show a \textbf{salient clustering pattern} where the entities with a similar type are clustered with each other. For example, the entities with the type \texttt{Person} are clustered.
Please note that the entities with the color red have location type, cyan is person, and violet is for the organization type.
While on the sub-figures labelled "a", the entities are scattered without any meaningful pattern in their adjacency. We believed that the size of the corpus could be responsible for this difference in performance. We based our assertion on the work conducted by \cite{antoniak2018evaluating}.\vspace{0.1cm}

\begin{figure}
\centering
 \subfigure[On enhanced corpus $(corpus^*)$.]{
   \includegraphics[width=0.45\columnwidth,height=4cm]{ 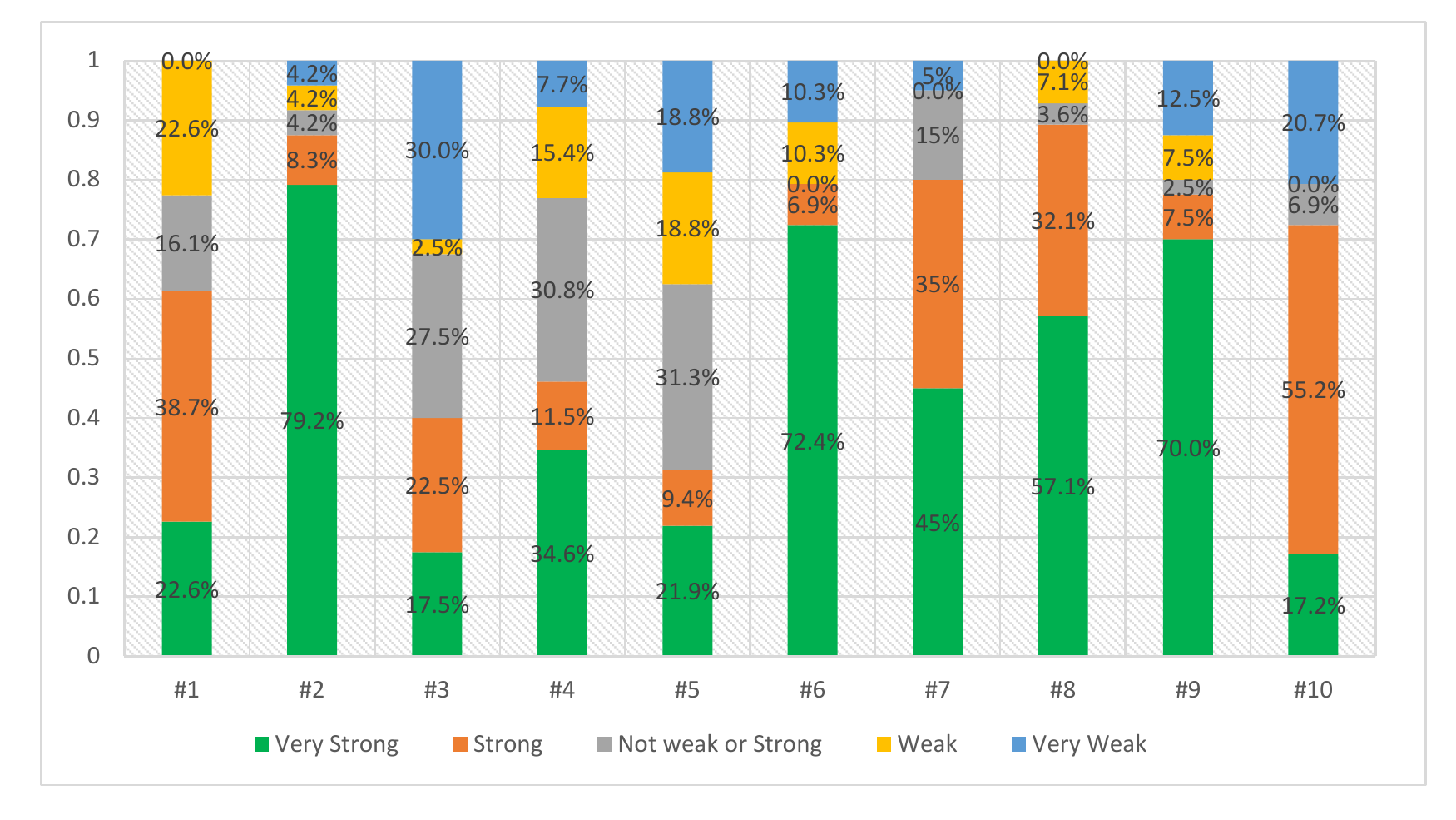}
   \label{fig:default_corpus}
	}
 \subfigure[On extended corpus $(corpus^+)$.]{
   \includegraphics[width=0.45\columnwidth,height=4cm]{ 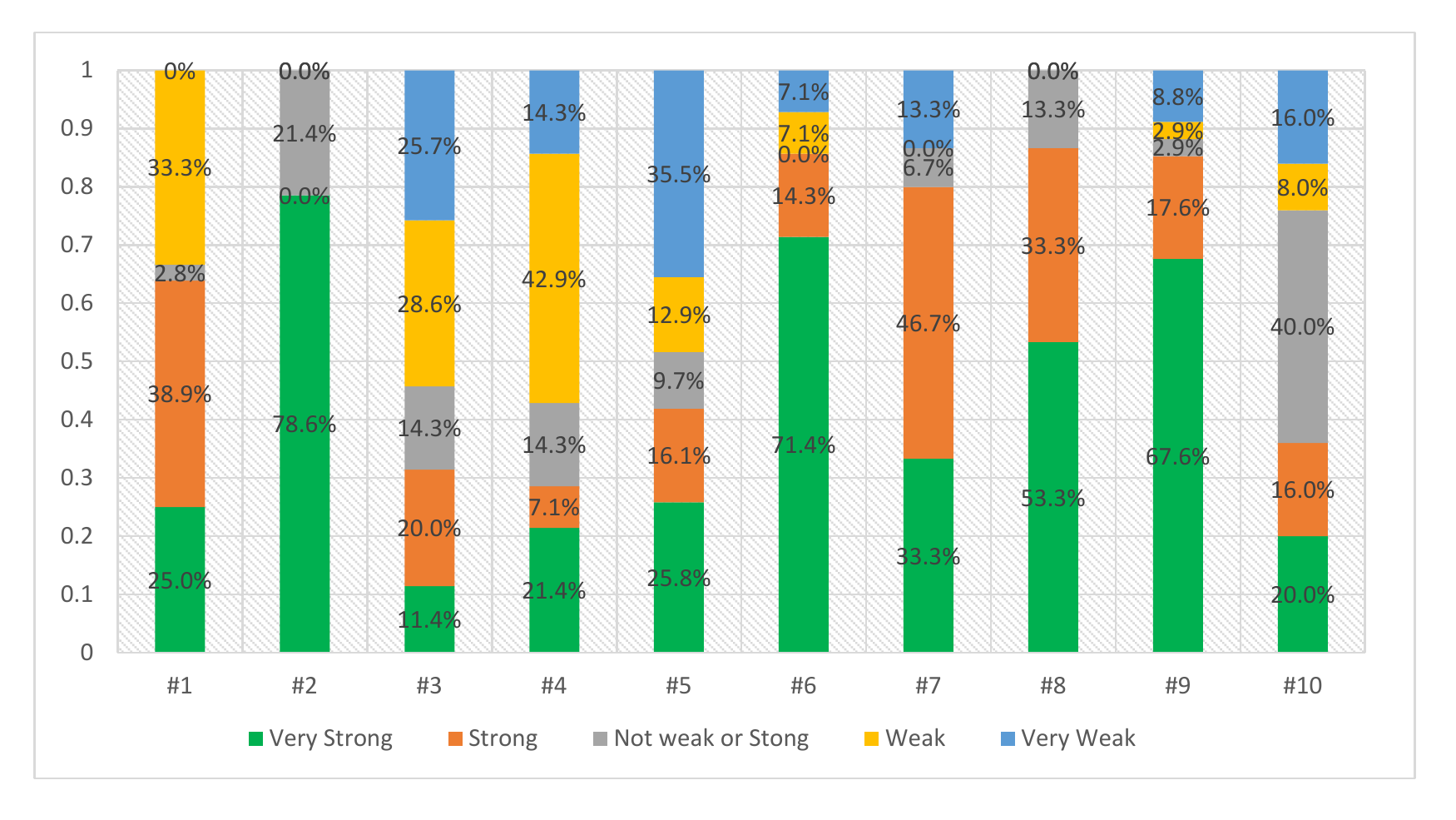}
    \label{fig:enhanced_corpus}
   }
 \caption{The results for evaluating the task of entailing associated entities using 10 human subjects.}
  \label{fig:human-judge}
\end{figure}

\noindent \textbf{Human Judge Approach.}
In this study, we selected ten people as our subject emerging entities. 
We generated an individual corpus for each subject. Then we captured the knowledge of each entity applying our approach on the corresponding corpus.
We prepared a Likert questionnaire listing the top-10 associated entities entailed by our model for $corpus^+$ and for $corpus^*$ respectively .
Then using this Likert, these judges (subjects) rate the associated entities based on their closeness to them.  
Figure \ref{fig:human-judge} shows the results of the evaluations for ten individual entities (each entity is identified by a number \#1 - \#10). 
Specifically, Figure \ref{fig:default_corpus} shows the evaluation on the extended corpus and Figure \ref{fig:enhanced_corpus} is concluded from the enhanced corpus.
In both of the charts, the green and orange colours respectively indicate the `very strong' and `strong' rates of the judge, meaning the captured entity is highly related to the subject.
On the contrary, the yellow and blue colours refer to respectively the `weak' and `very weak' ratings meaning the recognized entity is weakly associated with the subject.
The grey colour means that the judge could not recognize the entity (neutral).
Thus, with respect to the question as ``How well our approach can identify associated entities?'', we compare the ratings of `very strong', `strong', `weak' and `very weak') versus the ratings for unrecognized entities (neutral). 
On the enhanced corpus, on average, $>87\%$ of entities are recognizable, whereas it is $86\%$ on the extended corpus.
\begin{table}
\centering
\begin{scriptsize}
\begin{minipage}{.7\textwidth}
\caption{\scriptsize Comparison of  $corpus^+$ vs. $corpus^*$ . 
Performance \\ metrics:  P, R and F1 stand for respectively precision, \\ recall and F-score.}
\label{table:NP_metric_score}
\begin{tabular}{c|c|c|c||c|c|c}

\toprule
   & \multicolumn{3}{c|}{   $corpus^+$} &  \multicolumn{3}{|c|}{   $corpus^*$ } \\
\cline{2-7}
    Entity &   P &  R &  F1  &   P &  R &  F1  \\ 
\hline
\#1  &  .50 & .55 & .52 & .53 & .59  & .56 \\ 
\hline
\#2 & .86  & .50  &	.63  &	.89  &	.58  &	.70  \\
\hline
\#3 & .11  &	.61  &	.19  &	.45 &	.57  &	.50  \\
\hline
\#4 & .30  &	.45  &	.36  &	.50  &	.50  &	.50  \\
\hline
\#5 & .10  &	.47  &	.16  &	.59  &	.57  &	.58  \\
\hline
\#6 & .74  &	.56  &	.64  &	.76  &	.50  &	.60  \\
\hline
\#7 & .79  &	.43  &	.56  &	.85  &	.54  &	.66  \\
\hline
\#8 & .68  &	.59  &	.64  &	.87  &	.57  &	.69  \\
\hline
\#9 & .71  &	.70  &	.71  &	.83  &	.88  &	.86  \\
\hline
\#10 & .65  &	.52  &	.58  &	.68  &	.65  &	.66  \\
\midrule
 Average &  .54  &	 .54  &	 .50  &	 .70  & \textbf{.60}  &	 .63  \\
\bottomrule
\end{tabular}
\end{minipage}
\begin{minipage}{.7\textwidth}
\caption{\scriptsize Comparison of Google Knowledge Card with our approach \\ for 10 popular entities. Performance metrics: P, R and F1 \\ stand for respectively precision, recall and F-score. E. \\ and no. also stand for entities and number. }
\begin{tabular}{c|c|c|c|c}
\toprule
  \textbf{Entity} & \textbf{no. entailed E.}  & \textbf{no. Card E.}  & \textbf{R} & \textbf{P}\\ \toprule
\textbf{\#1} & 8  & 12 & .80 & .67                        \\ \hline
\textbf{\#2} & 7 & 7 & .70 & .100                       \\ \hline
\textbf{\#3} & 6 & 8 & .60 & .75                        \\ \hline
\textbf{\#4} & 4 & 7 & .40  & .57                        \\ \hline
\textbf{\#5} & 6 & 8 & .60 & .75                        \\ \hline
\textbf{\#6} & 5 & 6 & .50 & .83                        \\ \hline
\textbf{\#7} & 7 & 10 & .70 & .70                       \\ \hline
\textbf{\#8} & 6 & 8 & .60 & .75                        \\ \hline
\textbf{\#9} & 8 & 10 & .80 & .80                        \\ \hline
\textbf{\#10} & 5 & 9 & .50 & .56                        \\ 
\bottomrule

\end{tabular}

\label{tab:precision}
\end{minipage}
\end{scriptsize}
\end{table}
The next experimental question is ``How well does our model rank the associated entities''? To answer that, we compare the percentage of the entities rated as (`very strong', `strong') versus the entities rated as (`weak' and `very weak').
On the enhanced corpus, on average, 75\% of the associated entities have high closeness (`very strong', `strong') versus 17\% of them hold a weaker closeness.
Whereas on the extended corpus, on average, 62\% of the associated entities have high closeness (`very strong', `strong') while 26\% of them hold a weaker tie.
These observations show that although our approach is not a supervised approach, still the performance is fairly reasonable for both identifying associated entities and also ranking them based on the closeness criteria.
The other conclusion remark is that the enhanced corpus plays a significant role in the performance comparing to the extended corpus.
Also, we further re-represent our model performance using precision, recall, and F1 score metrics represented in Table \ref{table:NP_metric_score}.
On average the performance on $corpus^+$  reaches the precision (P), recall (R), and F1 score of  54\%, 54\%, and 50\%, while over the $corpus^*$, they are respectively 70\%, 60\%, and 63\%.
Thus, overall there is more than 5\% out-performance over $corpus^*$.\vspace{0.1cm}


\begin{figure}[hptb!]
\centering
 \subfigure[\texttt{Donald Trump}.]{
  \includegraphics[width=0.45\textwidth, height=7.5cm ]{ 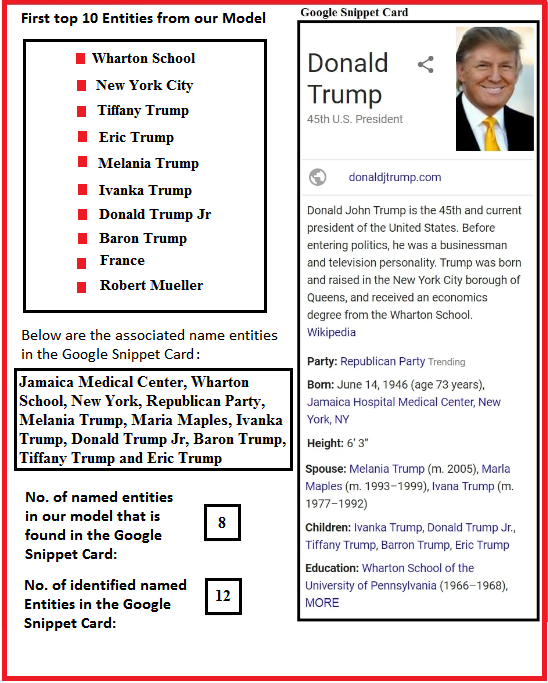}
   \label{fig:popular_entity_1}
	}
 \subfigure[\texttt{Lindsey Graham}.]{
   \includegraphics[width=0.45\textwidth, height=7.5cm]{ 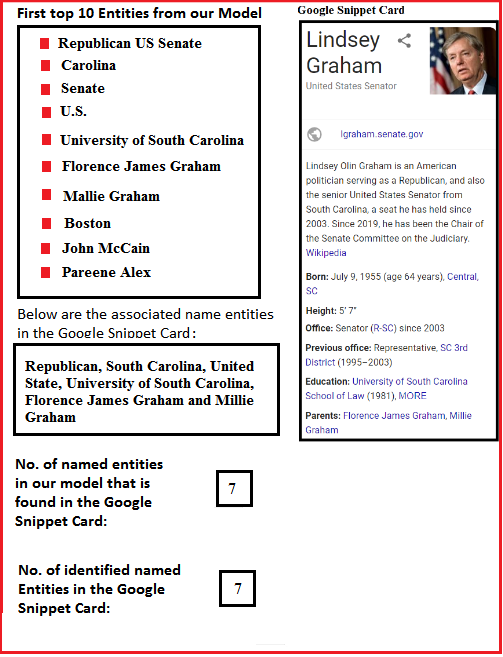}
    \label{fig:popular_entity_2}
   }
 \caption{The comparison of our entailment approach with Google Knowledge Card using popular entities as ground truth.}
 \label{fig:Google-card}
\end{figure}



\noindent\textbf{Comparison to Google Knowledge Card.}
In this experiment, we consider the ten popular entities from our testbed (\textit{precisely, the ones obtained with from enhanced corpus}), where they have a Google knowledge card. 
We compare our entailment model for the associated entities with the associated entities, which can be observed from their card. Two samples of such comparisons are shown in Figure \ref{fig:Google-card} where Figure \ref{fig:popular_entity_1} is about the entity \texttt{Donald Trump} and 
Figure \ref{fig:popular_entity_2} is about \texttt{Lindsey Graham}.
In total, Table \ref{tab:precision} summarizes the accuracy of entailing the associated entities. On average, we reach the precision of $62\%$ and recall of $73\%$ although our judgment is quite fair result based on fact that Google Knowledge Card only shows a summarized list of associated entities, therefore, there is the tendency of Google omitting some the entities our model captured. So, we assume that verifying with other knowledge graphs, perhaps can increase our precision and recall scores.\vspace{0.1cm}

\begin{figure}
\centering
   \includegraphics[width=0.8\textwidth, height=5cm]{ 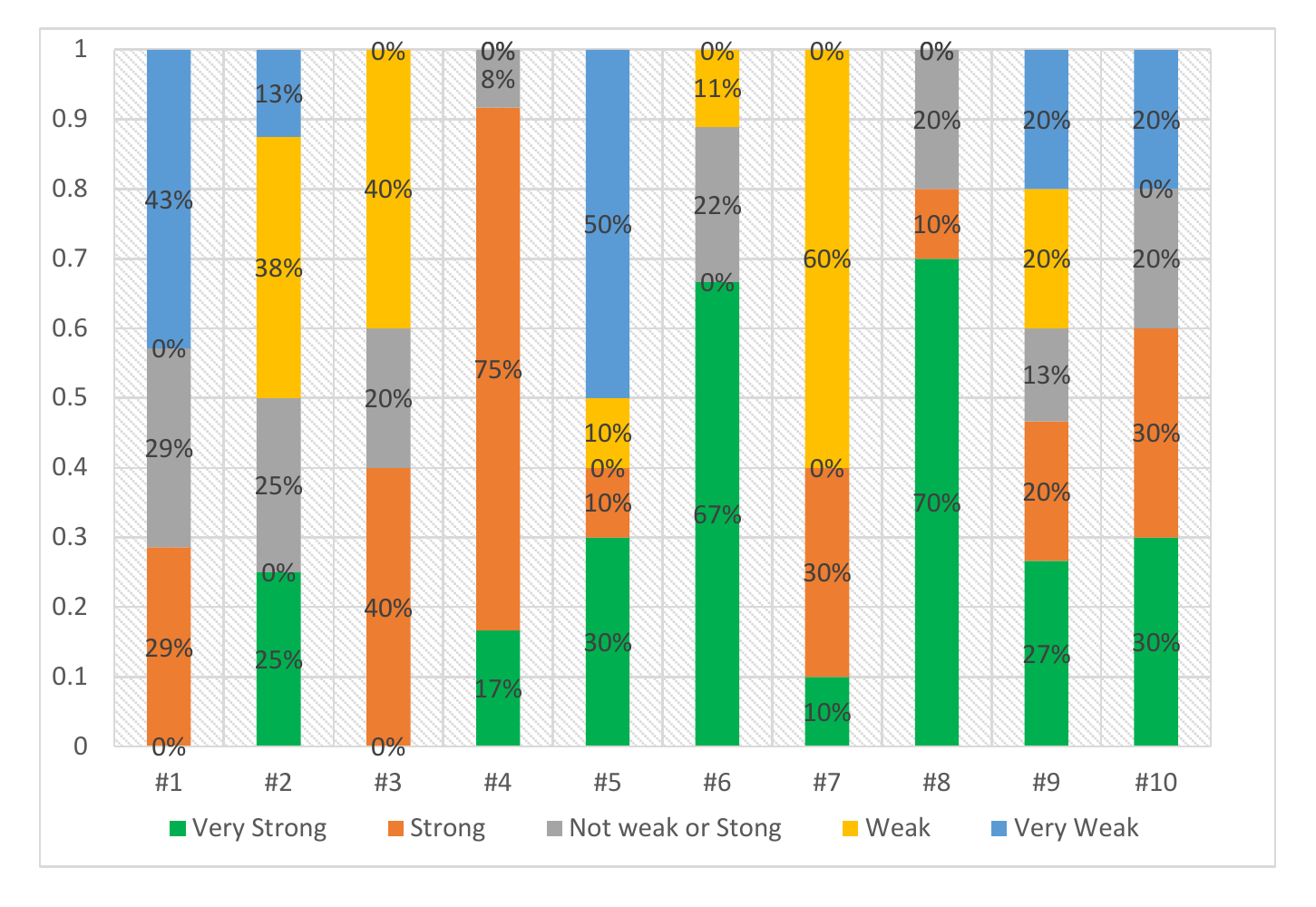}
   \caption{The judge result of ten emerging entities on entailing type using Likert questionnaire.}
   \label{fig:entity_typing}
\end{figure}

\noindent \textbf{Experimental Study of Entailing Type.}
The purpose of this experiment was to ascertain the performance of our model on assigning fine-grained types to a given emerging entity. To evaluate our model in this regard, we rely on the human judge approach, the judges were given Likert questionnaires where our Likert questionnaires that contain questions regarding the entailed types. Similar to the previous experiment, ten individual judges were participated to rate the appropriateness of their entailed types. The results of this study are illustrated in Figure \ref{fig:entity_typing}.
We consider the `neutral' (in grey colour) rating as the wrong entailment and the rest of the ratings as a recognizable type. Thus, with this respect, more than $87\%$ of the entailed types, are recognized by the judges.
Furthermore, on average, with $52\%$ of the ratings were either `strong' or `very strong'.
Although our approach for entailing type is a naive approach, the achieved results are promising. 
Also, the human judge is a subjective metric where we observe heterogeneity in their responses. 
Besides, we received much feedback from our subjects as this part seems more unaccustomed to them.
At last, please note that this analysis is done on the enhanced corpus $(corpus^*)$. \\
\vspace{-0.3cm}
\textbf{Conclusion and Future Plan.}
The emerging entities are those who have informative content on the Web while still, the encyclopedic knowledge graphs have not included them. 
In this paper, we presented unsupervised approaches for entailing the type and associated entities of the emerging entities.
The experimental study of our approach provides promising results.
Furthermore, we contribute to proposing an approach that takes the rank of the content into count for building up a textual corpus.
This approach is very effective where the content comes from a ranked list, such as search results. 
This research is the initial step of a longer agenda as we plan to employ supervised learning approaches in a combination of our current approach to enhance accuracy.
Furthermore, we plan to capture more fine-grained knowledge and the relation of the emerging entities.

\bibliographystyle{plain}
\bibliography{reference}
\end{document}